\documentclass[10pt,twocolumn, notitlepage]{article}
\usepackage{graphicx}
\usepackage{times}
\usepackage{natbib}
\begin{document}

\title{\Large\bf X-rays from young stars --  
        a summary of highlights from the \\ \textit{XMM-Newton Extended Survey of the Taurus Molecular Cloud (XEST)}\\ }

\vskip 0.4truecm
\author{{\bf M. G\"udel$^{1,2}$}\\  
\small $^1$Paul Scherrer Institut, W\"urenlingen \& Villigen, CH-5232 Villigen PSI, Switzerland; guedel@astro.phys.ethz.ch\\
\small $^2$Leiden Observatory, Leiden University, PO Box 9513, 2300 RA Leiden, The Netherlands
}

\date{\small Received 26 October 2007; accepted 30 October 2007}

\maketitle

\begin{abstract}
  \small The {\it XMM-Newton Extended Survey of the Taurus Molecular Cloud} (XEST) is a survey of the
  nearest large star-forming region, the Taurus Molecular Cloud (TMC), making use of all instruments on board 
  the {\it XMM-Newton} X-ray observatory. The
  survey, presently still growing, has provided unprecedented spectroscopic results from nearly every observed T Tauri star,
  and from $\approx$50\% of the studied brown dwarfs and protostars. The survey includes the first coherent statistical
  sample of high-resolution spectra of T Tauri stars, and is accompanied by an U-band/ultraviolet imaging photometric survey
  of the TMC. XEST led to the discovery of new, systematic  X-ray features not possible before with smaller samples, in particular 
  the {\it X-ray soft excess} in classical T Tauri stars and the {Two-Absorber X-ray} (TAX) spectra of jet-driving T Tauri stars.
  This paper summarizes highlights from XEST and  reviews the key role of this large project. 
\end{abstract}

\subsection*{1 Introduction}
Studies of star-forming regions have drawn a picture in \linebreak which cool, molecular gas,
contracting to stars with surrounding accretion disks, co-exists with high-energy
radiation. The latter is emitted by stellar plasma that is continuously
heated to temperatures of several million K. This radiation is most prominently
seen in the soft X-ray range and is conventionally attributed to the
presence of magnetic, coronal plasma.

In contrast to the solar case, high-energy radiation from forming stars has a profound influence
on the stellar environment. Extreme ultraviolet and X-ray photons ionize and heat dense gas in
accretion disks and the larger-scale molecular envelopes (e.g., Glassgold, Najita, \& Igea
2004). As a consequence, complex chemical networks may be driven. Further, ionized disk surface
layers make the gas accessible to magnetic fields, which then  induce the magnetorotational instability 
relevant for the accretion process (Balbus \& Hawley 1991). Ionized and magnetized environments
are also important for the disk-star accretion process through magnetic funnels, and for the acceleration
of bi-polar jets.

A deeper understanding of these physical mechanisms requires systematic studies of 
young stellar samples. The project described here has aimed at the nearest large star-form\-ation complex,
the Taurus Molecular Cloud (TMC), using all science instruments on board {\it XMM-Newton} (Jansen et al. 2001).

\subsection*{2 XEST and the Taurus Molecular Cloud}
The TMC is the closest and best-studied  large star formation region;
it has served as a test-bed for low-mass star-formation theory for decades.
The relevant region is quite large, however (some 10--15 degrees in diameter,
corres\-ponding to about 25--35~pc at a distance of 140~pc), making comprehensive studies
of the entire population difficult. 

The {\it XMM-Newton Extended Survey of the Taurus Mol\-ecular Cloud} (XEST) concentrates on the 
denser cloud areas that contain the bulk of the TMC stellar population. The emphasis is on 
a wide-field survey rather than on long exposures, although the total exposure time
has  now reached $\approx 1.3$~Ms. XEST goes systematically deeper than previous Taurus 
surveys by about an order of magnitude. A few exposures obtained by the
{\it Chandra X-Ray Observatory} have been added to the survey for completeness.

The concept, strategy, and databases of XEST are described in G\"udel et al. (2007a); the initial
science results have been published in 14 accompanying papers in a special feature of 
{\it Astronomy \& Astrophysics} (Vol. 468, 2007 June 3). XEST is a continuing survey, however. 
Following the initial 27 EPIC fields of view, each with a diameter of 30~arcmin, four 
additional exposures have been obtained, and at least
three further fields will become available in the archive.

\subsection*{3 The XEST stellar population}

The detection statistics of our EPIC  (Str\"uder et al. 2001; Turner et al. 2001)
X-ray survey is summarized in Table~\ref{statistics} separately for the 
{\it XMM-Newton} data, and for the combined {\it XMM-Newton} and {\it Chandra} fields (in parentheses).
The X-ray sample of detected classical TTS (CTTS) and weak-lined TTS (WTTS) is nearly complete for 
the surveyed areas as far as the population is known. Most of the few remaining, undetected objects are
considerably extincted and by implication X-ray absorbed.

\begin{table}[t]
\caption{XEST X-ray detection statistics of TMC stars and BDs. Numbers in parentheses include 
{\it Chandra} observations.\label{statistics} }
\begin{tabular}{lrrl}
\hline
\hline
Object type  & Members     &  Detections & Detection \\  
             & surveyed    &  	     & fraction (\%)  \\
\hline
Protostars	       & \phantom{1}20  \hfill (21) 	&     \phantom{12}8 \hfill (10)	& \phantom{1}40\% \hfill (48\%) \\
CTTS		       & \phantom{1}65  \hfill (70) 	&     \phantom{2}55 \hfill (60)	& \phantom{1}85\% \hfill (86\%) \\
WTTS		       & \phantom{1}50  \hfill (52) 	&     \phantom{2}49 \hfill (50)	& \phantom{1}98\% \hfill (96\%) \\
BD 		       & \phantom{1}16  \hfill (17) 	&     \phantom{12}8 \hfill (9)	& \phantom{1}50\% \hfill (53\%) \\
Herbig stars  	       & \phantom{15}2  \hfill {\ } 	&     \phantom{12}2 \hfill {\ }	&           100\% \hfill {\ }  \\
others/unident.        & \phantom{15}6  \hfill (7)  	&     \phantom{12}4 \hfill (5)	& \phantom{1}67\% \hfill (71\%) \\
\hline
Total		       &           159  \hfill (169)    &    126 \hfill (136)           & \phantom{1}79\% \hfill (80\%)	\\
\hline
\end{tabular}
\end{table}	

The Optical Monitor (Mason et al. 2001) has detected a total of 2148 point sources mostly in the U band, but 
partly also in ultraviolet (UV) bands (Audard et al. 2007); of these, 1893 objects have 2MASS counterparts, i.e., are
most likely to be stellar. Among them, 51 objects are known stellar/substellar members of TMC with X-ray counterparts. 
 
The Reflection Grating Spectrometers (den Herder et al. 2001) secured sufficiently good spectra of four CTTS, 
four WTTS, and one Herbig star, making this by far the largest coherent sample of high-resolution X-ray spectra of 
TTS available at the time (Telleschi et al. 2007b).

\subsection*{4 Accretion: UV and X-ray signatures}

Accretion is thought to proceed from the inner border of the CTTS accretion disk to the star via
magnetospheric accretion funnels, producing shocks near the photospheric layers; the shock-heated
gas is detected in large UV excess fluxes (e.g., Johns-Krull et al. 2000).
Although the distributions of bolometric (stellar) luminosities of the CTTS and WTTS samples agree with each other, 
the XEST OM survey indeed revealed, after  extinction corrections, a blue-/UV-excess in CTTS by up to 3 
magnitudes (Audard et al. 2007).

Variable accretion rates reflect in U-band and UV variability. A particularly striking example was seen
in an accreting brown dwarf (BD) in which the U band magnitude slowly increased during approximately 6 hours 
while the star remained entirely undetected in X-rays; Grosso et al. (2007b) concluded that the accretion 
rate increased by a factor of $\approx 3$ during that event.

\begin{figure}
\includegraphics[width=75mm]{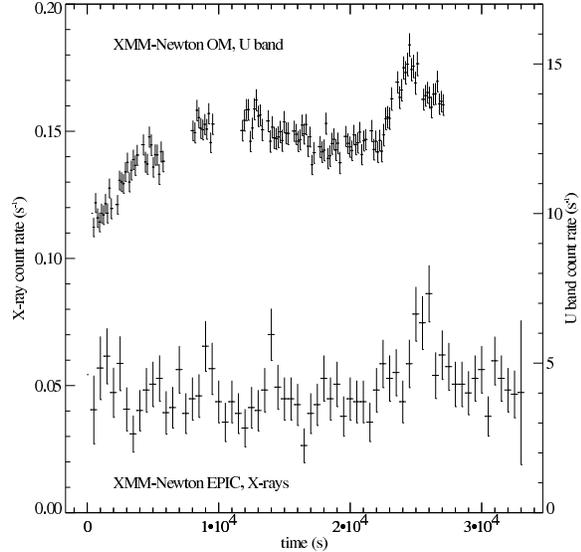}
\caption{U-band and X-ray light curves of the CTTS GK Tau. The correlated event
toward the end of the observation shows signatures of a flare, while the uncorrelated slow variation in
the U-band curve is most likely due to variable accretion (after Audard et al. 2007).}
\label{gktau}
\end{figure}

As this example shows, there is in general no correlation between X-ray and UV excess events, as is also 
evident in the large XEST collection of simultaneous UV and X-ray light curves; any correlated behavior
shows signatures of coronal flares, i.e., the UV event precedes the X-ray flare (Audard et al. 2007; 
Fig.~\ref{gktau}). As a consequence of accretion, however,  CTTS are much more UV variable  than WTTS.

\begin{figure}[b!]
\includegraphics[width=75mm]{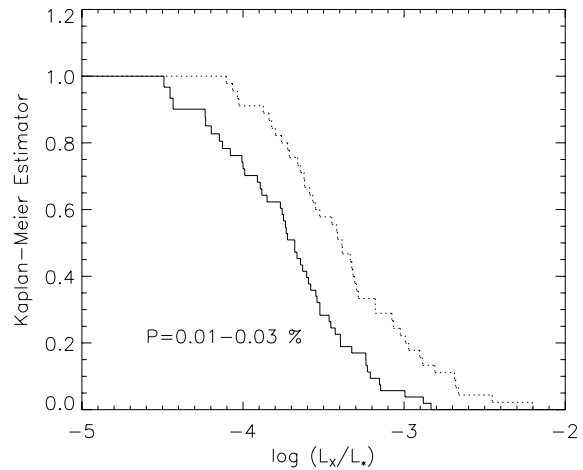}
\caption{Cumulative distributions (Kaplan-Meier estimator) for fractional X-ray luminosities, $L_{\rm X}/L_{\rm bol}$,
of classical (solid) and weak-line  (dotted) T Tauri stars in TMC (Telleschi et al. 2007a).}
\label{lxlbol}
\end{figure}

Are there {\it X-ray} signatures of accretion? XEST confir\-med a general X-ray deficiency by a factor
of  $\approx$2 for CTTS compared to WTTS (previously reported for other TTS, see Telleschi 
et al. 2007a for a summary; Fig.~\ref{lxlbol}), both in the $L_{\rm X}$ and $L_{\rm X}/L_{\rm bol}$ distributions 
while the 
distributions of $L_{\rm bol}$ for the two samples are indistinguishable. The cause of X-ray suppression is
not clear but appears to be related to a suppression of the coronal heating efficiency by inflowing 
cool, accreting material (Preibisch et al. 2005; Telleschi et al. 2007a). The distinction is less clear
among BDs; both accreting and non-accreting BDs show similar X-ray luminosity distributions, confirming 
the predominantly magnetic origin of the X-rays (Grosso et al. 2007a).

\subsection*{5 Discovery of the CTTS ``X-ray soft excess''}
Nearly free-falling gas will shock-heat  to maximum  temperatures 
$T_s = 8.6\times 10^5$~K~$[M/(0.5M_{\odot})] [R/(2R_{\odot})]^{-1}$  (Calvet \& Gullbring 1998). 
It is therefore conceivable that the softest  X-ray range reveals the high-$T$ tail of the 
shock emission measure. High electron densities up to $n_{\rm e}\approx 10^{13}$ cm$^{-3}$, measured in 
density-sensi\-tive X-ray lines of CTTS, have indeed been attributed to shocked gas (Kastner et al. 2002) although the 
XEST sample is ambiguous in this respect, showing  two accretors with low densities 
(AB Aur: Telleschi et al. 2007c; T Tau: G\"udel et al. 2007c).

On the other hand, the unprecedented, large XEST sample of  high-resolution RGS spectra 
revealed a new systematic not accessible before, namely an unusually high flux ratio between 
the O\,{\sc vii} and the   O\,{\sc viii} Ly$\alpha$ lines in CTTS. This excess flux, coined 
the {\it X-ray soft excess}, was first described by G\"udel (2006) and 
comprehensively discussed by Tell\-eschi et al. (2007b) and G\"udel et al. (2007c). 
\begin{figure}[t!]
\includegraphics[width=80mm]{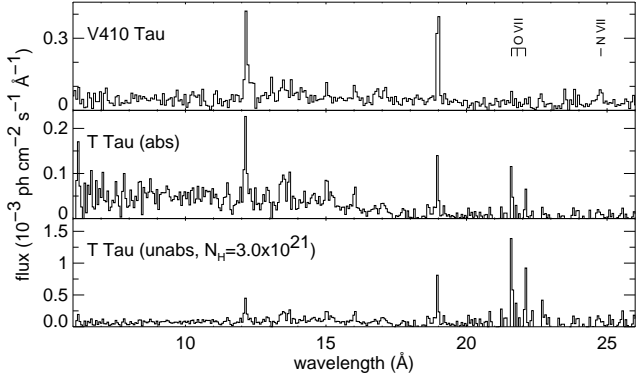}
\caption{Fluxed X-ray photon spectra of the WTTS V410~Tau, the CTTS T Tau, and the T Tau spectrum modeled 
         after removal of absorption (from top to bottom;  G\"udel \& Telleschi 2007).  }
\label{ttau}
\end{figure}

\begin{figure}[t!]
\includegraphics[width=80mm]{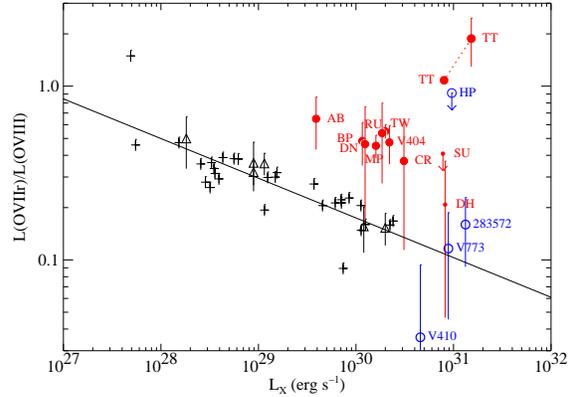}
\caption{The ratio between O\,{\sc vii} $r$ and O\,{\sc viii} Ly$\alpha$ luminosities vs. $L_{\rm X}$.  
Crosses mark MS stars, triangles solar analogs, filled (red) circles CTTS, and open (blue) circles WTTS. 
The solid line is a power-law fit to the MS stars (from G\"udel \& Telleschi 2007). }
\label{softexcess}
\end{figure}

The excessive O\,{\sc vii} flux is best illustrated for T Tau when compared to WTTS (or main-sequence [MS]
stars), and after removal of the photoelectric absorption in T Tau (Fig.~\ref{ttau}). The O\,{\sc vii} lines
are the strongest in the soft X-ray spectrum of T Tau, while they normally remain undetected in WTTS.
These lines are enhanced by typically a factor of 3--4 in CTTS compared to WTTS and equivalently active 
MS stars (Fig.~\ref{softexcess}). This is different from the UV excesses that are much stronger, and suggests a {\it 
relation with coronal activity} although {\it accretion is conditional for the X-ray soft excess.} 
The X-ray soft excess thus is {\it a new  accretion diagnostic for TTS discovered in the XEST project.}

\subsection*{6 The ``Two-Absorber X-Ray Sources''}
\begin{figure}[b!]
\centerline{\includegraphics[width=51mm,angle=-90]{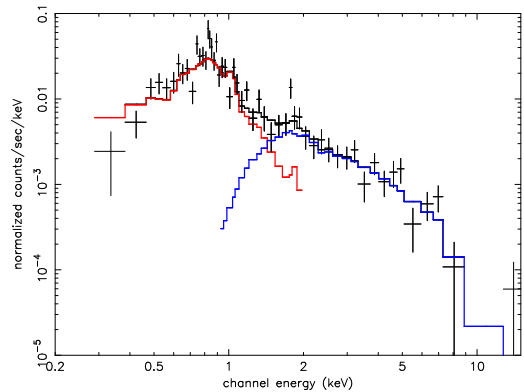}}
\caption{TAX spectrum of HN Tau (G\"udel et al. 2008, in prep).}
\label{tax}
\end{figure}

A new type of X-ray spectrum of CTTS was reported by G\"udel et al. (2005) and systematically
uncovered in several XEST CTTS (G\"udel et al. 2007b). The spectra of the 
{\it Two-Absorber X-ray} (TAX) sources show two unrelated  components 
subject to largely different photoelectric absorption. A soft component from a cool (few MK)
plasma, subject to very small absorption, is accompanied by a much harder component from
a hot (tens of MK) plasma, subject to about ten times higher absorption (Fig.~\ref{tax}). 
This latter absorption is, for standard interstellar gas-to-dust ratios, excessive if 
compared to the stellar visual extinction.

The TMC TAX sources (DG Tau, GV Tau, DP Tau, CW Tau - G\"udel et al. 2007b -- and
HN Tau -- Fig.~\ref{tax}) have in common i) a high accretion rate, accompanied
by a high mass outflow rate, and ii) the presence of ``micro-jets''. DG Tau's bi-polar
jet has meanwhile been identified in X-rays out to a distance of 5$^{\prime\prime}$ from the
star (G\"udel et al. 2008). 

The hard component flares in three TAX sources, suggesting a coronal/ma\-gnetospheric origin. 
The excessive absorption is attributed to 
massive accretion streams; they are, given their close distance to the star, dust depleted,
therefore not inducing excessive visual extinction. {\it The absorption of the hard component
thus provides unique signatures of accretion streams.}

The non-variable, weakly absorbed soft component cannot originate from the same regions; its close spectral
similarity with DG Tau's jet X-ray emission suggests an origin {\it at the base of the jets}. These
sources may thus  directly irradiate circumstellar disks from above.

\subsection*{7 A new abundance diagnostic?}
Depending on grain evolution in the circumstellar disk, certain elements may condense onto growing solids
and will not be accreted onto the star. If accretion feeds the X-ray source with new
material, an anomalous composition may result (e.g., Drake, Testa, \& Hartmann 2005). We found no
difference between CTTS and WTTS for abundance ratios between  Fe, Ne, and O determined from 
high-resolution spectroscopy. However, a dependence on spectral type or $T_{\rm eff}$  shows
up in that the Fe/Ne ratio is lower in later-type (K and M) stars than in G-type stars
(Telleschi et al. 2007b; G\"udel et al. 2007c).
This systematic in fact extends to active, ``X-ray saturated'' MS and post-MS stars as well
and is therefore not related to accretion. It has been confirmed for a larger  
sample of XEST EPIC spectra (Scelsi et al. 2007).

\subsection*{8 XEST: Successes and expansion}
XEST has  been among the first {\it XMM-Newton} ``Large Programs''. The relevance and 
importance of extended {\it XMM-Newton} programs is illustrated by systematic XEST findings that
were {\it not} feasible with the incoherent, smaller samples available before. XEST
has specifically succeeded in 
\begin{itemize}
\item the discovery of the accretion-related {\it X-ray soft excess}  of CTTS by securing the 
      largest uniform sample of  X-ray grating spectra of TTS simply not available before;
\item the systematic detection of several {\it two-absorber X-ray} sources now attributed to 
      jets;
\item the demonstration of new abundance systematics from  X-ray spectroscopy, 
      the Fe/Ne abundance ratio depending on spectral type but not on accretion properties;
\item a systematic study of the  (mostly accretion-related) 
      U-band/UV fluxes in the context of the (mostly coronal) X-rays 
      for a large sample of TTS, revealing the absence of correlated time behavior 
      except during flares;
\item the detection of an appreciable sample of X-ray emitting BDs, testifying to the 
      low detection limit of the survey.
\end{itemize}       
Many findings have remained unmentioned in this report, e.g., from studies of 
variability and flares, rotation, and the search for new TMC members.  Clearly, the concept
of a large program has been essential for XEST. A collection of restricted, non-uniform
small programs conducted by various teams with different goals and instrument settings will
not replace a uniformly conducted survey.

XEST is growing by several new exposures being added at the time of writing. Four additional 
exposures have been obtained, and at least three are becoming available in the archive. 
As far as these and future exposures are compatible with previous exposures of XEST, they 
will be uniformly reduced and analyzed, and then combined with the published survey. 
New results will be reported in upcoming papers.   

\begin{figure}[t!]
\includegraphics[width=81mm]{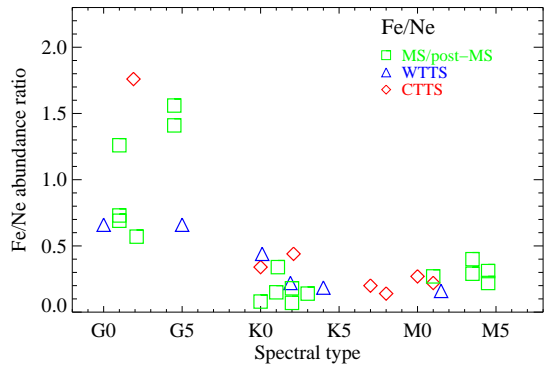}
\caption{Systematics in the Fe/Ne abundance ratio as a function of spectral type; stellar types
are as indicated by the symbols and colors. The ratios are relative to the solar photospheric ratio
 (after G\"udel et al. 2007c; Telleschi et al. 2007b). }
\label{rwaur}
\end{figure}

\noindent {\it Acknowledgments:}
We thank the International Space Science Institute (ISSI, Bern, Switzerland)  
for financial support of XEST. This research is based on observations obtained with 
{\it XMM-Newt\-on}, an ESA science mission with instruments and contributions directly 
funded by ESA member states and the USA (NASA).   


\subsection*{References}  
\noindent\small Audard, M., Briggs, K.R., Grosso, N., G\"udel, M., Scelsi, L., Bouvier, J., Telleschi, A. 2007: A\&A 468, 379\\
  Balbus, S.~A.,  Hawley J.~F.: 1991, ApJ 376, 214  \\
  Calvet, N., Gullbring, E.: 1998, ApJ 509, 802\\
  den Herder, J.~W., Brinkman, A.~C., Kahn, S.~M., et al.: 2001, A\&A 365, L7 \\
  Drake, J.~J., Testa, P., Hartmann, L.: 2005, ApJ 627, L149\\
  Glassgold, A.~E.,  Najita, J., Igea, J.: 2004,  ApJ 615, 972  \\
  Grosso, N., Audard A., Bouvier, J., Briggs, K.~R., G\"udel, M.: 2007b, A\&A 468, 557 \\  
  Grosso, N., Briggs, K.~R., G\"udel, M., et al.: 2007a, A\&A 468, 391   \\
  G\"udel, M. 2006, in High-Resolution X-Ray Spect\-ros\-copy; ed. G. Branduardi-Raymont,
             http:// www.mssl.ucl.ac.uk/$\sim$gbr/workshop2/; astro-ph/0609281\\
  G\"udel, M., Telleschi, A.: 2007, A\&A 474, L25  \\
  G\"udel, M., Briggs, K.~R., Arzner, K., et al. 2007a: A\&A 468, 353	 \\
  G\"udel, M., Skinner, S.~L., Audard, M., Briggs, K.~R., Cabrit, S.: 2008: A\&A, submitted \\ 
  G\"udel, M., Skinner, S.~L., Briggs, K.~R., et al.: 2005, ApJ 626, L53\\
  G\"udel, M., Skinner, S.~L., Mel'nikov, S.~Y., Audard, M., Telleschi, A., Briggs, K.~R.: 2007c, A\&A 468, 529     \\
  G\"udel, M., Telleschi, A., Audard, M., Skinner, S.~L., Briggs, K.~R., Palla, F.: 2007b, A\&A 468, 515   \\
  Jansen, F., Lumb, D., Altieri, B., et al.: 2001, A\&A 365, L1 \\
  Johns-Krull, C.M., Valenti, J.A., Linsky, J.L.: 2000, ApJ 539, 815\\
  Kastner, J.~H., Huenemoerder, D.~P., Schulz, N.~S., Canizares, C.~R., Weintraub, D.~A.: 2002, ApJ 567, 434\\
  Mason, K.~O., Breeveld, A., Much, R., et al.: 2001, A\&A 365, L36 \\
  Preibisch, T., Kim, Y.-C., Favata, F., et al.: 2005,  ApJS 160, 401 \\
  Scelsi, L., Maggio, A., Micela, G., Briggs, K.~R., G\"udel, M.: 2007, A\&A 473, 589\\
  Str\"uder, L., Briel, U.,  Dennerl, K., et al.: 2001, A\&A 365, L18\\
  Telleschi, A., G\"udel, M., Briggs, K.~R., Audard, M., Palla, F.: 2007a, A\&A 468, 425   \\    
  Telleschi, A., G\"udel, M., Briggs, K.~R., Audard, M., Scelsi, L.: 2007b, A\&A 468, 443\\     
  Telleschi, A., G\"udel, M., Briggs, K.~R., Skinner, S.~L., Audard, M., Franciosini, E.: 2007c, A\&A 468, 541 \\  
  Turner, M.~J.~L.,  Abbey, A., Arnaud, M., et al.: 2001, A\&A 365, L27

\end{document}